\documentclass[a4paper,10pt]{article}
\usepackage[fleqn]{amsmath}
\usepackage{amsfonts}
\usepackage{amssymb}
\title{Four Dimensional Supersymmetric Theories in Presence of a Boundary}
\author{   Mir Faizal$^1$ and Adel Awad$^{2,3}$\\
$^1$Department of Physics and Astronomy, \\  University of Waterloo,   Waterloo,\\
Ontario N2L 3G1, Canada\\
$^2$Center for Theoretical Physics, \\British University of Egypt\\
Sherouk City 11837, P.O. Box 43, Egypt\\
$^3$Department of Physics, Faculty of Science,\\
Ain Shams University, Abbassia, Cairo 11566, Egypt\\}
\date{}
\begin{document}

\maketitle

\begin{abstract}
In this paper, we study $\mathcal{N} =1$ supersymmetric theories in four dimensions in presence of a boundary.
We demonstrate that it is possible to preserve half the supersymmetry of the original theory by suitably 
modifying it in
presence of a boundary. This is done by adding new boundary terms to the original action, such that the
supersymmetric variation of the new terms exactly cancels the boundary terms generated by the supersymmetric 
transformation of the original bulk action.  We also analyze
the boundary projections of such supercharges used in such a theory.   
We study super-Yang-Mills theories in presence
of a boundary using these results. Finally, we     study the Born-Infeld action   
in presence of a boundary. We analyse the  boundary effects for 
the   Born-Infeld action coupled to a background dilaton  and an axion 
field. We also analyse the boundary effects for an non-abelian Born-Infeld  action. 
We explicitly construct the actions for these systems 
  in presence of a boundary. This action preserves half of the original supersymmetry. 
\end{abstract}

\section{Introduction}

The action for most renormalizable quantum field theories in four dimensions, including supersymmetric theories, is at most
quadratic in derivatives. So, variation of the action for such theories produces a bulk term as well as a total derivative term.
For manifolds without a boundary, the total derivative terms vanish due to the absence of a boundary. However, in presence of a boundary,
such total derivatives give rise to boundary contributions. The presence of a boundary breaks the translational invariance of the theory,
and this in turn breaks supersymmetry. In fact, supersymmetric variation of a supersymmetric action is known to be a total derivative. Thus,
in presence of a boundary, the supersymmetric variation of an action, which is supersymmetric in flat space, produces a nonvanishing boundary
term, this in turn breaks supersymmetry.

It is possible to restore supersymmetry on-shell by imposing some boundary conditions \cite{a}-\cite{b}.
There are various constraints generated from supersymmetry on the possible boundary conditions \cite{c}-\cite{g}.
However, this does not resolve the problem with the surface terms off-shell, since these boundary conditions are only imposed on the
on-shell fields,
and the supersymmetry is still broken off-shell. Since most supersymmetric theories are quantized using path integral formalism which uses
off-shell fields,
it is important to try to construct actions which preserve some supersymmetry off-shell.

Here we show that it is possible to construct an action which preserves half the original supersymmetry off-shell.
This can be done by modifying the original action through the addition
of boundary terms. The new boundary term added to the original action exactly cancel the boundary contribution generated from the supersymmetric
variation of the original bulk action. This procedure has been applied in three dimensions for $\mathcal{N} =1 $ supersymmetric theories\cite{1}.
The results thus obtained have been used for studying a system of multiple M2-branes ending on M5-brane \cite{4}-\cite{6}. As the gauge sector for the
action of  multiple M2-branes comprises of Chern-Simons theories, and the gauge transformation of Chern-Simons theories in presence of a boundary also
generates a boundary term, new boundary degrees of freedom had to be introduced on the boundary of the M2-branes. The gauge transformation of the action for these new boundary degrees of freedom exactly cancels the boundary contribution generated from the gauge transformation of the bulk action.
A system of   M2-branes intersecting with M5-branes have also been analysed
 in the supergravity regime using a  fuzzy funnel solution   \cite{d4}-\cite{d5}.

Apart from the M2-branes, the supersymmetric theory in presence of a boundary has also been used for analyzing non-anticommutativity in presence of a
boundary for a three dimensional theory with $\mathcal{N} =2$ supersymmetry \cite{8}. By suitably combining the boundary effects with
non-anticommutativity, a
three dimensional theory with $\mathcal{N} = 1/2$ supersymmetry has been constructed. In fact, the coupling of a three dimensional
super-Yang-Mills  theory  to background flux has been studied on a manifold with a boundary \cite{9}. The BRST symmetry for this system has also
been analyzed. However, all this work has been done in three dimensions.

  It may be noted that just like M2-branes can end on M5-branes, D3-branes can also end on other objects in string theory.  Such systems can be
  studied  using   fuzzy funnel \cite{f1}-\cite{f0}. In fact,
  a system of  D3-branes ending on other D3-branes
  have been analysed using  fuzzy funnel \cite{f5}. The   fuzzy funnel  have also been used
  to describe a system of D3-branes ending on D5-branes \cite{f4}, and  a system of D3-branes ending on  D7-branes  \cite{f2}.
 It would be interesting to apply to extend first develop a formalism for analysing four dimensional supersymmetry in presence
 of a boundaries, and then using such a formalism for studying D3-branes ending on various objects in string theory.
As the four dimensional super-Yang-Mills theory can be thought as a low energy limit of D3-brane action,
 we will analyze a four dimensional super-Yang-Mills theory in presence of a boundary.
 The construction of four dimensional supersymmetric theories in presence of a boundary can find
several other applications, and we are going to mention some of them in conclusion section of this paper.

The remaining paper is organized as follows. In section \ref{hgfa}, we discuss the general formalism for analysing $\mathcal{N} =1$
superfields in presence of a boundary, and construct a supersymmetric Lagrangian which preserves half of the original supersymmetry
in presence of a boundary.
In section \ref{hgfb},  we discuss the transformation of bulk and boundary superfields and supercharges
in presence of a boundary. In section \ref{hgfa1}, we apply this formalism to super-Yang-Mills theory. 
In section \ref{hgfa1b}, we will apply this formalism to Born-Infeld action. 
Finally, in section \ref{hgfb1},
we summarize our results and discuss some possible applications of the results of this paper.

 \section{Boundary Superfields}\label{hgfa}
 Let us start with a four dimensional  theory in  $\mathcal{N} =1$ superspace. This superspace  can  be parameterized by
   two supercharges,
$
 Q_{a} =-i \partial_{a} -(\gamma^\mu \partial_\mu\bar \theta)_{a}, $ and $
\bar Q_{a} = i\bar \partial_{a} + (\gamma^\mu \partial_\mu\theta)_{a},
$  which satisfy
 \begin{eqnarray}
\{Q_a, Q_b\} = 0,
&&
\{\bar Q_a, \bar Q_b\}= 0, \nonumber \\
 \{Q_a, \bar Q_b \} =  - 2i(\gamma^\mu \partial_\mu)_{ab}. &&
\end{eqnarray}
It is also  possible to define superderivates which commute with these generators of $\mathcal{N} =1$ supersymmetry,
$
 \{D_a, \bar Q_b \} =
\{D_a, \bar Q_b\} =
\{\bar D_a, \bar Q_b\} =
\{\bar D_a,  Q_b\} =0.
$
These superderivatives can be represented as
 $
  D_{a} = \partial_{a} + i(\gamma^\mu \partial_\mu\bar \theta)_{a},  $ and $
\bar D_{a}= \bar \partial_{a} + i(\gamma^\mu \partial_\mu\theta)_{a}
 $, and  satisfy
\begin{eqnarray}
\{D_a, D_b\} = 0,
&&
\{\bar D_a, \bar D_b\} =0, \nonumber \\
 \{D_a, \bar D_b \} =   2i(\gamma^\mu \partial_\mu )_{ab}. &&
\end{eqnarray}
Now we can write the Lagrangian for a supersymmetric theory with $\mathcal{N} =1$ supersymmetry as
\begin{eqnarray}
 \mathcal{L} &=&  D^2 \bar D^2 [\Phi (\theta, \bar\theta)]_{\theta = \bar\theta =0 }.
\end{eqnarray}

It may be noted that a linear combination of $\theta_a$ and $\bar \theta_a$ can be used to represent the
four dimensional supersymmetry,
 \begin{eqnarray}
\begin{pmatrix}\theta_{1a} \\ \theta_{2a} \end{pmatrix} &=&
 \begin{pmatrix}x_{11} & x_{12}\\
x_{21} & x_{22}  \end{pmatrix}\begin{pmatrix} \theta_{ a}\\ \bar \theta_{ a}\end{pmatrix}.
 \end{eqnarray}
 where $x_{ij}$ are  complex numbers such that, $x_{11}x_{22} - x_{12} x_{21} \neq 0$.
 We can write the original Lagrangian using these new coordinate as
 \begin{eqnarray}
  \mathcal{L} &=& D_1^2 D_2^2 \mathcal{J} [  \Phi (\theta_1, \theta_2)]_{\theta_1 = \theta_2 =0 },
 \end{eqnarray}
 where $\mathcal{J}$ is the Jacobian for transformation.
It is possible  absorb the Jacobian for transformation, using field redefinition, $\Phi (\theta_1, \theta_2) =
\mathcal{J}  \tilde \Phi (\theta_1, \theta_2) $, if $\tilde \Phi (\theta_1, \theta_2)$ is the original superfield.
We shall assume this to be this case and neglect the  numerical factor coming
from the Jacobian.
 Now we choose $x_{ij}$, such that in the new coordinates, the superderivatives take the form,
  \begin{eqnarray}
  D_{1a} &=& \partial_{1a} + (\gamma^\mu \theta_1)_a \partial_\mu,\nonumber \\
  D_{2a} &=& \partial_{2a} + (\gamma^\mu \theta_2)_a \partial_\mu,
 \end{eqnarray}
 and satisfy
  \begin{eqnarray}
 \{D_{1a}, D_{1b}\} = -2 \gamma_{ab}^{\mu}\partial_\mu, &&
  \{D_{2a}, D_{2b}\} =- 2 \gamma_{ab}^{\mu}\partial_\mu, \nonumber \\
  \{D_{1a}, D_{2b}\} = 0. &&
\end{eqnarray}
The generators of $\mathcal{N} =1$ supersymmetry corresponding to these superderivatives are given by
 \begin{eqnarray}
Q_{1a}  &=& \partial_{1a}  - (\gamma^\mu \theta_1)_a \partial_\mu,\nonumber \\
Q_{2a}  &=& \partial_{2a}  - (\gamma^\mu \theta_2)_a \partial_\mu,
 \end{eqnarray}
 and they also satisfy,
 \begin{eqnarray}
 \{Q_{1a}, Q_{1b}\} = 2 \gamma_{ab}^{\mu}\partial_\mu, &&
  \{Q_{2a}, Q_{2b}\} = 2 \gamma_{ab}^{\mu}\partial_\mu, \nonumber \\
  \{Q_{1a}, Q_{2b}\} = 0. &&
\end{eqnarray}
These supercharges also commute with
 these superderivatives,$\{Q_{1a}, D_{1b}\} = \{Q_{1a}, D_{2b}\} =0 $ and $
 \{Q_{2a}, D_{1b}\}= \{Q_{2a}, D_{2b}\} =0$.

We also  define,  $P_{\pm} = (1 \pm \gamma^3)/2$, so that $D_{1 \pm a} = (P_\pm)_{a}^{\;b} D_{1b}$ and $D_{2 \pm a} = (P_\pm)_{a}^{\;b} D_{2b}$.
We can also define
, and so,  $Q_{1 \pm a} = (P_\pm)_{a}^{\;b} Q_{1b}$ and $Q_{2 \pm a} = (P_\pm)_{a}^{\;b} Q_{2b}$  , we can write
the bulk charges $Q_{1a}$ and $Q_{2a}$ as
\begin{eqnarray}
  \epsilon^{1a} Q_{1a} &=&   \epsilon^{1a}( P_- + P_+) Q_{1a}  \nonumber \\ &=&
  \epsilon^{1+} Q_{1-} + \epsilon^{1-} Q_{1+},\nonumber \\
 \epsilon^{2a} Q_{2a} &=& \epsilon^{2a}( P_- + P_+) Q_{2a} \nonumber \\ &=&
  \epsilon^{2+} Q_{2-} + \epsilon^{2-} Q_{2+}.
\end{eqnarray}

Using the  super-derivative which commutes with the generator of $\mathcal{N} =1$ supersymmetry,
we can write a Lagrangian for a supersymmetric theory with $\mathcal{N} =1$ supersymmetry as
\begin{eqnarray}
 \mathcal{L}  &=& D^2_2 D_1^2 [\Phi (\theta_1, \theta_2)]_{\theta_1 = \theta_2 =0} \nonumber \\
 &= & D_2^2 [r_2(\theta_2)]_{\theta_2=0} \nonumber \\
 &= & D^2_1 [r_1(\theta_1)]_{\theta_1=0},
\end{eqnarray}
where the $\mathcal{N} =1$ superfield has been decomposed as
\begin{eqnarray}
  \Phi ( \theta_1, \theta_2)&=& p_1 ( \theta_1) + q_1 (\theta_1) \theta_2 + r_1 (\theta_1) \theta_2^2 \nonumber \\ &=&
 p_2 ( \theta_2) + q_2 (\theta_2) \theta_1 + r_2 (\theta_2) \theta_1^2.
\end{eqnarray}
It may be noted that  $p_1 ( \theta_1), p_2 ( \theta_2), q_1 (\theta_1), q_2 (\theta_2), r_1
(\theta_1), r_2 (\theta_2) $ are   superfields
in their own right depending only on $\theta_2$ or $\theta_1$.
Under  the supersymmetric
transformations generated by $Q_{1a}$ and $Q_{2a}$, they transform as
\begin{eqnarray}
   \epsilon^{1a} Q_{1a} p_1 (\theta_1) &=& \epsilon^{1a} q_{1a}(\theta_1), \nonumber \\
\epsilon^{1a} Q_{1a} q_{1a}(\theta_1) &=&  -\epsilon_{1a} r_1 (\theta_1) + (\gamma^\mu\epsilon_1)_a \partial_a p_1(\theta_1), \nonumber \\
\epsilon^{1a} Q_{1a} r_1 (\theta_1) &=& \epsilon^{1a} (\gamma^\mu \partial_\mu)_a^b q_{1b}(\theta_1), \nonumber \\
   \epsilon^{2a} Q_{2a} p_2 (\theta_2) &=& \epsilon^{2a} q_{2a}(\theta_2), \nonumber \\
\epsilon^{2a} Q_{2a} q_{2a}(\theta_2) &=&  -\epsilon_{2a} r_2(\theta_2) + (\gamma^\mu\epsilon_2)_a \partial_a p_2(\theta_2), \nonumber \\
\epsilon^{2a} Q_{2a} r_2 (\theta_2) &=& \epsilon^{2a} (\gamma^\mu \partial_\mu)_a^b q_{2b}(\theta_2).
\end{eqnarray}
Thus, under these supersymmetric transformations generated by $Q_{1a}$  this Lagrangian transforms
as $\epsilon^{1a} Q_{1a} \mathcal{L}= - \partial_\mu (\gamma^\mu\epsilon^1 q_1 (\theta_1))$,
and under these supersymmetric transformations generated by $Q_{2a}$  this Lagrangian transforms
as $\epsilon^{2a} Q_{2a} \mathcal{L}= - \partial_\mu (\gamma^\mu\epsilon^2 q_2 (\theta_2))$. So, the action is invariant under the supersymmetric
transformations generated by $Q_{1a}$ and $Q_{2a}$, in absence of a boundary, $\epsilon^{1a} Q_{1a} \mathcal{L} = \epsilon^{2a} Q_{2a} \mathcal{L}
=0$.
However, in presence of a boundary, the  supersymmetric
transformations generated by $Q_{1a}$ and $Q_{2a}$ produce boundary terms. Thus, if we assume that a boundary exists at $x_3 =0$,
then  the supersymmetric transformations of
the Lagrangian can be written as   $
\epsilon^{1a} Q_{1a}  \mathcal{L} = - \gamma^3 \partial_3 ( \epsilon^{1a} q_{1a} (\theta_1)) $ and  $
 \epsilon^{2a} Q_{2a}  \mathcal{L} = - \gamma^3 \partial_3 ( \epsilon^{2a} q_{2a} (\theta_2)) $.
The presence of these boundary terms will  breaks the supersymmetry of the resultant theory.

We can perverse half the supersymmetry of the resultant theory by either adding or subtracting a boundary term to the
original Lagrangian. Now if  $\mathcal{L}_{1b}$ and $\mathcal{L}_{2b}$
is the boundary term added or subtracted from the bulk Lagrangian with $\mathcal{N} =1$ supersymmetry, then we have
\begin{eqnarray}
\epsilon^1 Q_1 [  \mathcal{L}  \pm  \mathcal{L}_{1b}] &=&
 \pm 2  \partial_3 \epsilon^{1\pm}  q_{1\mp } (\theta_1), \nonumber
 \\
\epsilon^2 Q_2 [ \mathcal{L}  \pm  \mathcal{L}_{2b}]  &=&
  \pm 2  \partial_3 \epsilon^{2\pm}  q_{2\mp} (\theta_2),
\end{eqnarray}
where $q_{1\pm }  (\theta_1) = (1 \pm  \gamma^3) q_1 (\theta_1)/2   $ 
and $q_{2\pm }(\theta_2) = (1 \pm  \gamma^3) q_2(\theta_2)/2$. 
Hence, the Lagrangian $\mathcal{L} \pm \mathcal{L}_{1b}$    preserves the
supersymmetry generated by
$\epsilon^{1\mp} Q_{1\pm}$,
and
the the Lagrangian $\mathcal{L} \pm \mathcal{L}_{2b}$    preserves the
supersymmetry generated by
$\epsilon^{2\mp} Q_{2\pm}$. It is not possible to simultaneously preserve both the supersymmetry generated by  $\epsilon^{1-} Q_{1+}$ and
$\epsilon^{1+} Q_{1-}$, or  $\epsilon^{2-} Q_{2+}$ and
$\epsilon^{2+} Q_{2-}$,
in the presence of a boundary.
However, in the presence of a boundary, we can construct the Lagrangian
which preserves the supersymmetry
generated by $\epsilon^{1\mp} Q_{1\pm}$ and $\epsilon^{2\mp} Q_{2\pm}$ as
\begin{eqnarray}
 \mathcal{L}^{1- 2 -} &=& (D^2_1 - \partial_3) (D^2_2 - \partial_3)
[\Phi (\theta_1, \theta_2)]_{\theta_1 = \theta_2 =0}, \nonumber \\
 \mathcal{L}^{1- 2 +} &=& (D^2_1 - \partial_3) (D^2_2+ \partial_3)
[\Phi (\theta_1, \theta_2)]_{\theta_1 = \theta_2 =0}, \nonumber \\
 \mathcal{L}^{1+ 2 -} &=& (D^2_1 + \partial_3) (D^2_2 - \partial_3)
[\Phi (\theta_1, \theta_2)]_{\theta_1 = \theta_2 =0}, \nonumber \\
 \mathcal{L}^{1+ 2 + } &=& (D^2_1+ \partial_3) (D^2_2 + \partial_3)
[\Phi (\theta_1, \theta_2)]_{\theta_1 = \theta_2 =0}.
\end{eqnarray}
It may be noted that this Lagrangian preserves only half of the supersymmetry of the original Lagrangian.
This is because if we preserve the supersymmetry corresponding to $\epsilon^{1\mp} Q_{1\pm}$ and $\epsilon^{2\mp} Q_{2\pm}$,
then we will break the supersymmetry corresponding to $\epsilon^{1\mp} Q_{1\mp}$ and $\epsilon^{2\mp} Q_{2\mp}$.

It may be noted that half the on-shell supersymmetry could also be preserved by 
 using suitable boundary conditions. In fact, these on-shell boundary 
 conditions can be motivated from this off-shell formalism. This is because the supersymmetric 
 transformation of the original Lagrangian are given by 
 \begin{eqnarray}
\epsilon^1 Q_1 [  \mathcal{L}  \pm  \mathcal{L}_{1b}] &=&
 \pm 2   {\epsilon^{1\pm}}'  q'_{1\mp } (\theta_1), \nonumber
 \\
\epsilon^2 Q_2 [ \mathcal{L}  \pm  \mathcal{L}_{2b}]  &=&
  \pm 2   {\epsilon^{2\pm}}'  q'_{2\mp} (\theta_2),
\end{eqnarray}
 where  $'$ means the quantity is evaluated at the boundary. 
 As  
the supersymmetric transformation of $\mathcal{L}  \pm  \mathcal{L}_{1b}$ do  not 
generate   ${\epsilon^{1\mp}}'  q'_{1\pm } (\theta_1)$,  and the 
supersymmetric transformation of $\mathcal{L}  \pm  \mathcal{L}_{2b}$ do  not 
generate  ${\epsilon^{2\ mp}}'  q'_{2\pm} (\theta_2)$, this Lagrangian is invariant 
under half the off-shell supersymmetry. 
However, half of the on-shell supersymmetry could also be preserved by 
imposing the  following boundary conditions on the original Lagrangian,   
 \begin{eqnarray}
   q'_{1-} (\theta_1) =0,   &&
   q'_{2-} (\theta_2) =0,   \\
   q'_{1- } (\theta_1) =0,  &&   
   q'_{2+} (\theta_2) =0,   \\ 
   q'_{1+} (\theta_1) =0, &&  
   q'_{2-} (\theta_2) =0,  \\
   q'_{1+} (\theta_1) =0,  &&
   q'_{2+} (\theta_2) =0.  
  \end{eqnarray}
As these terms would vanish on-shell by the imposition of the boundary conditions, 
this Lagrangian is also invariant under half of the on-shell 
supersymmetry of the original Lagrangian. 
As we have defined 
$ \Phi ( \theta_1, \theta_2)= p_1 ( \theta_1) + q_1 (\theta_1) \theta_2 + r_1 (\theta_1) \theta_2^2 =
 p_2 ( \theta_2) + q_2 (\theta_2) \theta_1 + r_2 (\theta_2) \theta_1^2, 
$,  we can write 
\begin{eqnarray}
 q_{1a}(\theta_1) = [D_{2a} \Phi (\theta_1, \theta_2)]_{\theta_2 =0}, && 
  q_{2a}(\theta_2) = [D_{1a} \Phi (\theta_1, \theta_2)]_{\theta_1 =0}.
\end{eqnarray}
Thus, on the boundary we can write 
 \begin{eqnarray}
   q'_{1a-} (\theta_1) &=& P^b_{a -}[D_{1b} \Phi (\theta_1, \theta_2)]'_{\theta_1 =0},   \\
   q'_{2a-} (\theta_2) &=& P^b_{a -}[D_{2b} \Phi (\theta_1, \theta_2)]'_{\theta_2 =0},   \\
   q'_{1a- } (\theta_1) &=&P^b_{a-}[D_{1b} \Phi (\theta_1, \theta_2)]'_{\theta_1 =0}, \\   
   q'_{2a+} (\theta_2) &=&P^b_{a +}[D_{2b} \Phi (\theta_1, \theta_2)]'_{\theta_2 =0},    \\ 
   q'_{1a+} (\theta_1) &=&P^b_{a +}[D_{1b} \Phi (\theta_1, \theta_2)]'_{\theta_1 =0},  \\  
   q'_{2a-} (\theta_2) &=&P^b_{a -}[D_{2b} \Phi (\theta_1, \theta_2)]'_{\theta_2 =0},   \\
   q'_{1a+} (\theta_1) &=&P^b_{a +}[D_{1b} \Phi (\theta_1, \theta_2)]'_{\theta_1 =0},  \\
   q'_{2a+} (\theta_2) &=&P^b_{a +}[D_{2b} \Phi (\theta_1, \theta_2)]'_{\theta_2 =0},   
  \end{eqnarray}
    where the projection operator is defined by 
    $P^b_{a \pm} = [\delta^b_a\pm  (\gamma^3)_{a }^b] /2$ 
    and $'$ indicates that only the boundary values are considered. 
Now half  the on-shell supersymmetric can also be 
 preserved by imposing the following boundary conditions on the superfield,
 \begin{eqnarray}
 P^b_{a +}{[D_{1b} \Phi (\theta_1, \theta_2)]}'_{\theta_1 =0} =0, &&
 P^b_{a +}{[D_{2b} \Phi (\theta_1, \theta_2)]}'_{\theta_2 =0}=0,  \\
 P^b_{a +}{[D_{1b} \Phi (\theta_1, \theta_2)]}'_{\theta_1 =0}=0, &&   
 P^b_{a -}{[D_{2b} \Phi (\theta_1, \theta_2)]}'_{\theta_2 =0}=0,   \\ 
 P^b_{a -}{[D_{1b} \Phi (\theta_1, \theta_2)]}'_{\theta_1 =0}=0, &&  
 P^b_{a +}{[D_{2b} \Phi (\theta_1, \theta_2)]}'_{\theta_2 =0}=0,  \\
 P^b_{a -}{[D_{1b} \Phi (\theta_1, \theta_2)]}'_{\theta_1 =0}=0, &&
 P^b_{a -}{[D_{2b} \Phi (\theta_1, \theta_2)]}'_{\theta_2 =0}=0.  
  \end{eqnarray}
  It is important to note that these boundary conditions are invariant under half the generators of supersymmetry.  
 This is needed for these boundary conditions to hold under supersymmetric transformations. 
  Even though we can preserve half the   on-shell supersymmetric of the original Lagrangian by imposing 
 these boundary conditions, the advantage of using the present formalism is that it also preserves half 
 of the off-shell supersymmetry.
 
\section{Transformation of Boundary Fields}\label{hgfb}
In this section, we will analyse the decomposition of the supercharges for a four dimensional theory
with  $\mathcal{N} =1$ supersymmetry.
We can write the bulk supercharges as
$
  \epsilon^{1a} Q_{1a}
= \epsilon^{1+} Q_{1-} + \epsilon^{1-} Q_{1+}, $ and $
 \epsilon^{2a} Q_{2a}
= \epsilon^{2+} Q_{2-} + \epsilon^{2-} Q_{2+}.
$
Furthermore, the the bulk supercharges $Q_{1\pm},    Q_{2\pm},  $  are related to
boundary supercharges $Q'_{1\pm},   Q'_{2\pm },  $  as
\begin{eqnarray}
 Q_{1-} = Q'_{1- }+ \theta_{1-} \partial_3, &&
 Q_{1+} = Q'_{1+} - \theta_{1+}\partial_3,  \nonumber \\
  Q_{2-}= Q'_{2- }+ \theta_{2-} \partial_3, &&
 Q_{2+} = Q'_{2+} - \theta_{2+}\partial_3,
\end{eqnarray}
The boundary
supercharges given by
\begin{eqnarray}
  Q'_{1+} = \partial_{1+} - \gamma^s \theta_{1-} \partial_s, &&
   Q'_{1-} = \partial_{1-} - \gamma^s \theta_{1+} \partial_s,\nonumber \\
  Q'_{2+} = \partial_{2+} - \gamma^s \theta_{2-} \partial_s, &&
   Q'_{2-} = \partial_{2-} - \gamma^s \theta_{2+} \partial_s,
\end{eqnarray}
where $s$
is the
index for the coordinates along the boundary,  i.e.,  the case $\mu = 3$ has been excluded for a boundary fixed at $x_3$.
Now by definition $Q_{1\pm },   Q_{2\pm}, $ are the generators of the
half supersymmetry of the bulk fields and
$Q'_{1\pm}, , Q'_{2\pm},  $ are the standard
supersymmetry of the boundary fields. We can now express the boundary superfields in terms of bulk superfields as follows,
\begin{eqnarray}
 Q'_{1-} &=& \exp ( + \theta_{1+} \theta_{1-} \partial_3)
 Q_{1-} \exp ( - \theta_{1+} \theta_{1-} \partial_3),
\nonumber \\
 Q'_{1+} &=& \exp ( -  \theta_{1-} \theta_{1+} \partial_3)
 Q_{1+}\exp ( +  \theta_{1-} \theta_{1+} \partial_3),
 \nonumber \\
  Q'_{2-} &=& \exp ( + \theta_{2+} \theta_{2-} \partial_3)
 Q_{2-}\exp ( - \theta_{2+} \theta_{2-} \partial_3),
\nonumber \\
 Q'_{2+} &=& \exp ( -  \theta_{2-} \theta_{2+} \partial_3)
 Q_{2+}\exp ( +  \theta_{2-} \theta_{2+} \partial_3).
\end{eqnarray}
The original superfield also gets decomposed as follows,
\begin{eqnarray}
 \Phi &=&  \exp ( +  \theta_{2-} \theta_{2+} \partial_3)\exp ( +  \theta_{1-} \theta_{1+} \partial_3) \Phi'_{2+ 1+},
 \nonumber \\
  \Phi &=&  \exp ( +  \theta_{2-} \theta_{2+} \partial_3) \exp ( - \theta_{1+} \theta_{1-} \partial_3) \Phi'_{2+1-},
  \nonumber \\
   \Phi &=&  \exp ( - \theta_{2+} \theta_{2-} \partial_3)\exp ( +  \theta_{1-} \theta_{1+} \partial_3) \Phi'_{2- 1+},
 \nonumber \\
  \Phi &=&  \exp ( - \theta_{2+} \theta_{2-} \partial_3) \exp ( - \theta_{1+} \theta_{1-} \partial_3) \Phi'_{2-1-},
\end{eqnarray}
where $\Phi'_{2+ 1+},\Phi'_{2+ 1-}, \Phi'_{2- 1+},\Phi'_{2- 1-}, $ decompose into boundary superfields,
\begin{eqnarray}
  \epsilon^{1-} Q_{1+} \Phi &=& \exp ( +  \theta_{2-} \theta_{2+} \partial_3)\exp ( +  \theta_{1-} \theta_{1+} \partial_3)
 {\epsilon^{1-}}' Q_{1+}' \Phi'_{2+ 1+}, \nonumber \\
  \epsilon^{1+} Q_{1-} \Phi &=& \exp ( +  \theta_{2-} \theta_{2+} \partial_3) \exp ( - \theta_{1+} \theta_{1-} \partial_3)
  {\epsilon^{1+}}' Q_{1-}' \Phi'_{2+ 1-}, \nonumber \\
  \epsilon^{2-} Q_{2+} \Phi &=& \exp ( +  \theta_{2-} \theta_{2+} \partial_3)\exp ( +  \theta_{1-} \theta_{1+} \partial_3)
 {\epsilon^{2-}}' Q_{2+}' \Phi'_{2+ 1+}, \nonumber \\
  \epsilon^{2+} Q_{2-} \Phi &=& \exp ( - \theta_{2+} \theta_{2-} \partial_3)\exp ( +  \theta_{1-} \theta_{1+} \partial_3)
 {\epsilon^{2+}}' Q_{2-}' \Phi'_{2- 1+}, \nonumber \\
   \epsilon^{1-} Q_{1+} \Phi &=& \exp ( - \theta_{2+} \theta_{2-} \partial_3)\exp ( +  \theta_{1-} \theta_{1+} \partial_3)
 {\epsilon^{1-}}' Q_{1+}' \Phi'_{2-1+}, \nonumber \\
  \epsilon^{1+} Q_{1-} \Phi &=& \exp ( - \theta_{2+} \theta_{2-} \partial_3) \exp ( - \theta_{1+} \theta_{1-} \partial_3)
  {\epsilon^{1+}}' Q_{1-}' \Phi'_{2- 1-}, \nonumber \\
  \epsilon^{2-} Q_{2+} \Phi &=& \exp ( +  \theta_{2-} \theta_{2+} \partial_3) \exp ( - \theta_{1+} \theta_{1-} \partial_3)
 {\epsilon^{2-}}' Q_{2+}' \Phi'_{2+ 1-}, \nonumber \\
  \epsilon^{2+} Q_{2-} \Phi &=& \exp ( - \theta_{2+} \theta_{2-} \partial_3) \exp ( - \theta_{1+} \theta_{1-} \partial_3)
 {\epsilon^{2+}}' Q_{2-}' \Phi'_{2- 1-}.
\end{eqnarray}

Now we will   analyse the superalgebra for the four dimensional theory with
  $\mathcal{N}= 1$ supersymmetric theory,
 in the presence of a boundary. The non-vanishing part of the superalgebra is given by
\begin{eqnarray}
 \{Q_{1+ a}, Q_{1+ b}\} = 2 (\gamma_{ab}^{s}P_+)\partial_s   ,
 &&  \{D_{1+a}, D_{1+b}\} =- 2 (\gamma_{ab}^{s}P_+)\partial_s  , \nonumber \\
 \{Q_{1- a}, Q_{1- b}\} = 2 (\gamma_{ab}^{s}P_-)\partial_s   ,
 &&  \{D_{1-a}, D_{1-b}\} =- 2 (\gamma_{ab}^{s}P_-)\partial_s   , \nonumber \\
 \{Q_{1+a}, Q_{1-b}\} = -2 (P_{-})_{ab}\partial_3   ,
 &&  \{D_{1+a}, D_{1-b}\} = 2 (P_-)_{ab}\partial_3   , \nonumber \\
 \{Q_{2  + a}, Q_{2  + b}\} = 2 (\gamma_{ab}^{s}P_+)\partial_s   ,
 &&  \{D_{2  +a}, D_{2  +b}\} =- 2 (\gamma_{ab}^{s}P_+)\partial_s  , \nonumber \\
 \{Q_{2  - a}, Q_{2  - b}\} = 2 (\gamma_{ab}^{s}P_-)\partial_s   ,
 &&  \{D_{2  -a}, D_{2  -b}\} =- 2 (\gamma_{ab}^{s}P_-)\partial_s   , \nonumber \\
 \{Q_{2  +a}, Q_{2  -b}\} = -2 (P_{-})_{ab}\partial_3   ,
 &&  \{D_{2  +a}, D_{2  -b}\} = 2 (P_-)_{ab}\partial_3 .
\end{eqnarray}
It may be noted that $\{Q_{1\pm}, Q_{2\pm}\} = \{D_{1\pm}, D_{2\pm}\} =0$, and
$\{Q_{1\pm}, D_{2\pm}\} = \{Q_{1\pm}, D_{1\pm}\} =\{Q_{2\pm}, D_{2\pm}\} = \{Q_{2\pm}, D_{1\pm}\} =0$.
So, we have
\begin{eqnarray}
 D_{1-a}D_{1+b} = (P_-)_{ab} (\partial_3 -D_1^2), &&
 D_{1+a}D_{1-b} =  -(P_-)_{ab} (\partial_3 + D_1^2),
  \nonumber \\
   D_{2-a}D_{2+b} = (P_-)_{ab} (\partial_3 -D_2^2), &&
 D_{2+a}D_{2-b} = -(P_-)_{ab} (\partial_3 + D_2^2).
\end{eqnarray}
Now contracting these equation and using $(P_-)_a^a =1$, we
obtain the following result,
\begin{eqnarray}
D_1^2 + \partial_3 =  D_{1+}D_{1-}, && D_2^2 + \partial_3 =D_{2+}D_{2-}, \label{a1} \\
D_1^2 - \partial_3 = D_{1-}D_{1+},&& D_2^2 - \partial_3 =D_{2-}D_{2+}.\label{a4}
\end{eqnarray}
Thus, we can see how the  Lagrangian with the measure preserves
 the right amount of supersymmetry on the boundary, because we can write
 \begin{eqnarray}
   \mathcal{L}^{1+ 2+} &=&  D_{2+}D_{2-}D_{1+}D_{1-}
 [  \Phi ]_{\theta_1 = \theta_2=0},
\nonumber \\
   \mathcal{L}^{1- 2-} &=&  D_{2-}D_{2+}D_{1-}D_{1+}
 [  \Phi ]_{\theta_1 = \theta_2=0},
\nonumber \\
   \mathcal{L}^{1+ 2-} &=&  D_{2+}D_{2-}D_{1-}D_{1+}
 [  \Phi ]_{\theta_1 = \theta_2=0},
\nonumber \\
   \mathcal{L}^{1- 2+} &=& D_{2-}D_{2+}D_{1+}D_{1-}
 [  \Phi ]_{\theta_1 = \theta_2=0}.
 \end{eqnarray}
We can write it in terms of boundary superfields as
\begin{eqnarray}
 \mathcal{L}^{1+ 2+}  &=&
- D_{2+}'D_{1+}' [ \Omega'_{1- 2-} ]_{\theta_{1-} = \theta_{2-} =0},
\nonumber \\
 \mathcal{L}^{1- 2-}  &=&
- D_{2-}'D_{1-}' [ \Omega'_{1+ 2+} ]_{\theta_{1+} = \theta_{2+} =0},
\nonumber \\
 \mathcal{L}^{1+ 2-}  &=&
- D_{2-}'D_{1+}' [ \Omega'_{1- 2+} ]_{\theta_{1-} = \theta_{2+} =0},
\nonumber \\
 \mathcal{L}^{1- 2+}  &=&
- D_{2+}'D_{1-}' [ \Omega'_{1+ 2-} ]_{\theta_{1+} = \theta_{2--} =0},
\end{eqnarray}
where $'$ means the quantity is evaluated at the boundary and
\begin{eqnarray}
 \Omega'_{1- 2-} &=&  D'_{2-}D'_{1-} [\Phi'_{1- 2-} ]_{\theta_{1-} = \theta_{2-} =0},
 \nonumber \\
  \Omega'_{1+ 2+} &=& D'_{2+}D'_{1+}[\Phi'_{1+ 2+} ]_{\theta_{1+} = \theta_{2+} =0},
\nonumber \\
 \Omega'_{1- 2+} &=&  D'_{2+}D'_{1-}[\Phi'_{1- 2+} ]_{\theta_{1-} = \theta_{2+} =0},
 \nonumber \\
  \Omega'_{1+2-} &=&   D'_{2-}D'_{1+}[\Phi'_{1+2-} ]_{\theta_{1+} = \theta_{2-} =0},
 \end{eqnarray}
The boundary measure   only  contains $D'_{2\pm}D'_{1\pm}$.
So, on the boundary ${\epsilon^{1\pm}}' Q_{1\mp}'$ and ${\epsilon^{2\pm}}'
 Q_{2\mp}'$ act as independent supercharges.
Thus, we can add a boundary Lagrangian to the original theory, which will still preserve half the supersymmetry
of the original theory,
\begin{eqnarray}
 \mathcal{L}_t &=& \mathcal{L} + \mathcal{L}_b,
\end{eqnarray}
where $\mathcal{L}_t$ is the total Lagrangian for the bulk and the boundary theory, $\mathcal{L}$ is the Lagrangian for the
original   theory and $\mathcal{L}_b$ is the Lagrangian for the boundary theory.
Thus, we can add the following terms to Lagrangian
\begin{eqnarray}
 \mathcal{L}^{1+ 2+}  &=&
- D_{2+}'D_{1+}' [ \Omega'_{1- 2-} +  \omega'_{1- 2-}  ]_{\theta_{1-} = \theta_{2-} =0},
\nonumber \\
 \mathcal{L}^{1- 2-}  &=&
- D_{2-}'D_{1-}' [ \Omega'_{1+ 2+}  +\omega'_{1+ 2+}  ]_{\theta_{1+} = \theta_{2+} =0},
\nonumber \\
 \mathcal{L}^{1+ 2-}  &=&
- D_{2-}'D_{1+}' [ \Omega'_{1- 2+} + \omega'_{1- 2+}]_{\theta_{1-} = \theta_{2+} =0},
\nonumber \\
 \mathcal{L}^{1- 2+}  &=&
- D_{2+}'D_{1-}' [ \Omega'_{1+ 2-}  + \omega'_{1+ 2-}  ]_{\theta_{1+} = \theta_{2--} =0}.
\end{eqnarray}
Here $\omega'_{1\pm 2 \pm}$   are only defined on the boundary,
\begin{eqnarray}
 \omega'_{1- 2-} &=&  D'_{2-}D'_{1-} [\lambda'_{1- 2-} ]_{\theta_{1-} = \theta_{2-} =0},
 \nonumber \\
  \omega'_{1+ 2+} &=& D'_{2+}D'_{1+}[\lambda'_{1+ 2+} ]_{\theta_{1+} = \theta_{2+} =0},
\nonumber \\
 \omega'_{1- 2+} &=&  D'_{2+}D'_{1-}[\lambda'_{1- 2+} ]_{\theta_{1-} = \theta_{2+} =0},
 \nonumber \\
  \omega'_{1+2-} &=&   D'_{2-}D'_{1+}[\lambda'_{1+2-} ]_{\theta_{1+} = \theta_{2-} =0},
 \end{eqnarray}
where $\lambda'_{1\pm 2\pm}$ can be  purely boundary Lagrangian. We can take a suitable
gauge invariant coupling between this purely boundary fields and the bulk fields.
It may be noted that  on the boundary only
the
supersymmetry generated by  ${\epsilon^{1\pm}}' Q_{1\mp}'$ and ${\epsilon^{2\pm}}'
 Q_{2\mp}'$ is preserved.

 \section{Super-Yang-Mills Theory}\label{hgfa1}
 In this section,
  we will write the action for super-Yang-Mills theory as using a vector field $V^AT_A$, where $T_A$ are the generators
 of the gauge symmetry, $[T_A, T_B] = i f^C_{AB} T_C$.
We can write
the Lagrangian for the super-Yang-Mills theory as using a vector superfield $ V = V^A T_A $,
 \begin{eqnarray}
  \mathcal{L} &=& D^2  [W^a W_a]_{\theta= 0} + \bar D^2 [\bar W^a \bar W_a]_{\bar \theta = 0} \nonumber \\
  && + \bar D^2  D^2  [\mathcal{V}(\Phi,\bar \Phi)
  + \bar \Phi e^{V} \Phi ]_{\theta = \bar\theta =0 }
  \nonumber \\ &=&   D^2 \bar D^2 [\Box^{-1}   D^2   W^a W_a     +
 + \Box^{-1} \bar D^2    \bar W^a \bar W_a    \nonumber \\
  &&  + \mathcal{V}(\Phi,\bar \Phi)
  + \bar \Phi e^{V} \Phi ]_{\theta = \bar\theta =0 },
 \end{eqnarray}
 where the superfield strengths are given by $W_a = - i \bar D^2 (e^{-V} D_a
 e^V)/4$ and $\bar W_a = - i  D^2 (e^{-V} \bar D_a
 e^V)/4$. Here the potential $\mathcal{V}(\Phi,\bar \Phi)$ is a function of $\Phi$ and $\bar \Phi$.
 Even though this action looks like a non-local action, the component action in the bulk will be a local action. This
 is because it is another way of writing a local action. It may be noted, as  we were only interested in analysing
 the amount of supersymmetry preserved, we will did not need the explicit form of
super-Yang-Mills action in real superfields. It may be noted that even though the expression for it would  involve an complicated expression
 containing the non-local operator, the component action would be  local. This is because it can be transformed back
 into the local action. However, it is not clear if the resultant boundary action is local or not, as it cannot be transformed into
 any local action. So, we will express this Lagrangian into an alternative formalism, and in that formalism we will be
 able to obtain a local action for the super-Yang-Mills theory even in presence of a  boundary.

 The   gauge transformations of the superfield  $V$ transforms are given by
 $e^V \to e^{i \bar \Lambda} e^V e^{-i  \Lambda} $, where $\Lambda$
 and $\bar \Lambda$ are chiral and anti-chiral gauge parameters.
So, it is possible to write a covariant derivative which transforms under gauge transformation as
 $ \nabla_a = e^{-V} D_a e^V \to e^{i \Lambda}\nabla_a e^{- i \Lambda}   $ and  $ \bar \nabla_a  =
 \bar D_a \to e^{i \Lambda}\nabla_a e^{- i \Lambda}   $,
 and another   covariant derivative which transforms under gauge transformation as
 $ \tilde \nabla_a  = D_a \to e^{i \bar \Lambda}\tilde \nabla_a e^{- i \bar\Lambda}   $
 and  $ \tilde {\bar{\nabla}}_a = e^V \bar D_a e^{-V} \to e^{i \bar \Lambda}\tilde{\bar{\nabla}}_a e^{- i \bar\Lambda}$.
 However, it  is also possible to define another
  covariant derivative which transforms under a real gauge  parameter $u$ as
   $  \nabla_a \to u  \nabla_a u^{-1}   $ and
  $  \bar{\nabla}_a \to u   \bar{\nabla}_a u^{-1}   $ \cite{1001}.
 Now  we can express this covariant derivative in terms of   $\theta_{1a} $ and $\theta_{2a} $ rather than
 $\theta_{1a} $ and $\theta_{2a} $.  We can
 absorb the Jacobian using field redefinition,
 and then use the modified  measure on the boundary.
 However, it would be more convenient to express the original  covariant derivative
 in terms of the real spinor  superfield and then work out the modification by the boundary theory. So,
 we define two gauge valued   spinor superfields $\Gamma_{1a}= \Gamma_{1a}^A  (\theta_1) T_A$ and $\Gamma_{2a}
 = \Gamma_{2a}^A  (\theta_2) T_A $, and use them to construct
  gauge covariant derivatives for matter fields $\Phi (\theta_1, \theta_2 )$ and $\bar \Phi(\theta_1, \theta_2 )$,
\begin{eqnarray}
 \nabla_{1a} \Phi =D_{1a}\Phi -i \Gamma_{1a} \Phi, &&
 \nabla_{2a} \Phi = D_{2a}\Phi -i \Gamma_{2a} \Phi,  \nonumber \\
 \nabla_{1a} \bar \Phi =D_{1a}\bar \Phi +i \bar \Phi\Gamma_{1a} , &&
 \nabla_{2a} \bar  \Phi = D_{2a}\bar \Phi +i \bar \Phi\Gamma_{2a}.
\end{eqnarray}
These fields transform   under the
  gauge transformation as,
$
\Gamma_{1a}  \to  u\nabla_{1a} u^{-1},$  and $
 \Gamma_{2a}  \to  u\nabla_{2a} u^{-1},
$ and so the covariant derivatives transform  as
$\nabla_{1a} \to u \nabla_{1a} u^{-1}$ and $ u \nabla_{2a} u^{-1}$.
It may be noted if we define $\nabla_a$ and $\bar\nabla_{a}$ as a linear combination of
$\nabla_{1a}  $ and $ u  $, then we will get to correct transformation
for the original covariant derivatives. This is because
$\nabla_a = x_{11}\nabla_{1a} + x_{12} \nabla_{2a} \to u [x_{11}\nabla_{1a} + x_{12} \nabla_{2a} ]u^{-1}
= u\nabla_a u^{-1}$ and
$\bar \nabla_a = x_{22}\nabla_{2a} + x_{21} \nabla_{1a}
\to u [x_{22}\nabla_{2a} + x_{21} \nabla_{1a}  ]u^{-1}
= u\bar\nabla_a u^{-1}$, where $x_{ij}$ are complex numbers.
We can also construct the field strengths as follows,
\begin{eqnarray}
W_{1a} &=&  \frac{1}{2} D^b_1 D_{1a} \Gamma_{1b} - \frac{i}{2}  \{\Gamma^b_1, D_{1b} \Gamma_{1a}\}
- \frac{1}{6} [ \Gamma^b_1 ,
\{ \Gamma_{1b}, \Gamma_{1a}\}],\nonumber \\
W_{2a} &=&  \frac{1}{2} D^b_2 D_{2a} \Gamma_{2b} - \frac{i}{2}  \{\Gamma^b_2, D_{2b} \Gamma_{2a}\}
- \frac{1}{6} [ \Gamma^2_1 ,
\{ \Gamma_{2b}, \Gamma_{2a}\}].
\end{eqnarray}
Now these field strengths transform as $
W_{1a}  \to  u W_{1a} u^{-1},$  and $
W_{2a}  \to  u W_{2a} u^{-1}.
$
Now we can write the action for super-Yang-Mills theory as
\begin{eqnarray}
 \mathcal{L} &=& D^2_1 D^2_2 [ \nabla^a \Phi  \bar \nabla_a  \bar  \Phi
 + \mathcal{V} [\Phi, \bar \Phi]]_{\theta_1 = \theta_2 =0}\nonumber \\&&
+ D^2_1 [W^a_1W_{1a}]_{\theta_1 =0} + D^2_2
[W^a_2W_{2a}]_{\theta_2 =0},
\end{eqnarray}
where $\mathcal{V} [\Phi, \bar \Phi]$ is a potential term
 which is given by product of superfields $\Phi$ and $\bar\Phi$.

 Now we can write the Lagrangian for super-Yang-Mills theory which preserves various supercharges as follows,
\begin{eqnarray}
 \mathcal{L}^{1- 2 -} &=& (D^2_1 - \partial_3) (D^2_2 - \partial_3)
[ \nabla^a \Phi  \bar \nabla_a  \bar  \Phi
 + \mathcal{V} [\Phi, \bar \Phi]]_{\theta_1 = \theta_2 =0}
  \nonumber \\&&+ (D^2_1 - \partial_3) [W^a_1W_{1a}]_{\theta_1 =0} + (D^2_2 - \partial_3) [ W^a_2W_{2a}]_{\theta_2 =0}
 , \nonumber \\
 \mathcal{L}^{1- 2 +} &=& (D^2_1 - \partial_3) (D^2_2+ \partial_3)
[ \nabla^a \Phi  \bar \nabla_a  \bar  \Phi
 + \mathcal{V} [\Phi, \bar \Phi]]_{\theta_1 = \theta_2 =0}
 \nonumber \\&& +(D^2_1 - \partial_3) [W^a_1W_{1a}]_{\theta_1 =0} + (D^2_2+ \partial_3)[ W^a_2W_{2a}]_{\theta_2 =0}
 , \nonumber \\
 \mathcal{L}^{1+ 2 -} &=& (D^2_1 + \partial_3) (D^2_2 - \partial_3)
[ \nabla^a \Phi  \bar \nabla_a  \bar  \Phi
 + \mathcal{V} [\Phi, \bar \Phi]]_{\theta_1 = \theta_2 =0}
  \nonumber \\&&+(D^2_1 + \partial_3) [W^a_1W_{1a}]_{\theta_1 =0} + (D^2_2 - \partial_3)[ W^a_2W_{2a}]_{\theta_2 =0}
 , \nonumber \\
 \mathcal{L}^{1+ 2 + } &=& (D^2_1+ \partial_3) (D^2_2 + \partial_3)
[ \nabla^a \Phi  \bar \nabla_a  \bar  \Phi
 + \mathcal{V} [\Phi, \bar \Phi]]_{\theta_1 = \theta_2 =0}
  \nonumber \\&&+  (D^2_1+ \partial_3) [W^a_1W_{1a}]_{\theta_1 =0} + (D^2_2 + \partial_3) [ W^a_2W_{2a}]_{\theta_2 =0}
 .
\end{eqnarray}
This result can be also be written as
 \begin{eqnarray}
   \mathcal{L}^{1+ 2+} &=&  D_{2+}D_{2-}D_{1+}D_{1-}
 [  \nabla^a \Phi  \bar \nabla_a  \bar  \Phi
 + \mathcal{V} [\Phi, \bar \Phi]]_{\theta_1 = \theta_2=0}
  \nonumber \\&&+ D_{1+}D_{1-} [W^a_1W_{1a}]_{\theta_1 =0} + D_{2+}D_{2-} [ W^a_2W_{2a}]_{\theta_2 =0}
 ,
\nonumber \\
   \mathcal{L}^{1- 2-} &=&  D_{2-}D_{2+}D_{1-}D_{1+}
 [  \nabla^a \Phi  \bar \nabla_a  \bar  \Phi
 + \mathcal{V} [\Phi, \bar \Phi]]_{\theta_1 = \theta_2=0}
  \nonumber \\&&+ D_{1-}D_{1+} [W^a_1W_{1a}]_{\theta_1 =0} + D_{2-}D_{2+} [ W^a_2W_{2a}]_{\theta_2 =0}
 ,
\nonumber \\
   \mathcal{L}^{1+ 2-} &=&  D_{2+}D_{2-}D_{1-}D_{1+}
 [ \nabla^a \Phi  \bar \nabla_a  \bar  \Phi
 + \mathcal{V} [\Phi, \bar \Phi]]_{\theta_1 = \theta_2=0}
  \nonumber \\&&+ D_{1-}D_{1+} [W^a_1W_{1a}]_{\theta_1 =0} + D_{2+}D_{2-} [ W^a_2W_{2a}]_{\theta_2 =0}
 ,
\nonumber \\
   \mathcal{L}^{1- 2+} &=& D_{2-}D_{2+}D_{1+}D_{1-}
 [ \nabla^a \Phi  \bar \nabla_a  \bar  \Phi
 + \mathcal{V} [\Phi, \bar \Phi] ]_{\theta_1 = \theta_2=0}
  \nonumber \\&&+ D_{1+}D_{1-} [W^a_1W_{1a}]_{\theta_1 =0} + D_{2-}D_{2+} [ W^a_2W_{2a}]_{\theta_2 =0}.
 \end{eqnarray}
  It is thus transparent that this modified Lagrangian only preserves half the original
 supersymmetry.
 
 \section{The Born-Infeld Action} \label{hgfa1b}
 
 This action can be thought as a low energy action generated from 
 the Born-Infeld action, which 
 is the action for D3-branes. It is possible to write the full  
  Born-Infeld  Lagrangian in superspace  \cite{1b, 2b, 3b, 4b}. 
  The abelian Born-Infeld can be written as \cite{5b}
  \begin{eqnarray}
   S  = \frac{1}{(2 \pi \alpha')^2} \int d^4 x \sqrt{ - \rm{det} (\eta_{\mu\nu} + (2 \pi \alpha') F_{\mu\nu})}, 
  \end{eqnarray}
where $F_{\mu\nu} = \partial_\mu A_\nu - \partial_\nu A_\mu$. It is also possible to express the abelian 
Born-Infeld action using complex bosonic variables, 
\begin{eqnarray}
\omega = \alpha + i \beta, && \bar \omega = \alpha - i \beta, \nonumber \\
 \alpha = \frac{1}{4} F^{\mu\nu}F_{\mu\nu}, && \beta = \frac{1}{4} F^{\mu\nu}\tilde{F}_{\mu\nu}, 
\end{eqnarray}
where $\tilde{F}_{\mu\nu}$ is defined as $\tilde{F}_{\mu\nu} = \epsilon^{\mu\nu \tau \rho} F_{\mu\nu}/2 $. 
So, the   abelian Born-Infeld Lagrangian can be written as   \cite{1b}
\begin{eqnarray}
 \int d^4 x  \mathcal{L}  =  \int d^4 x  -\frac{1}{2}(\omega + \bar \omega ) +
 (2 \pi \alpha')^2 \omega \bar \omega \mathcal{B}(\omega, \bar \omega).
\end{eqnarray}
The function $\mathcal{B}(\omega, \bar \omega)$ can be expressed as 
  \begin{eqnarray}
  B (\omega , \bar \omega ) =   \left[1- \frac{(2 \pi\alpha')^2}{2}\omega_+
 + \sqrt{ 1 + (2 \pi\alpha')^2\omega_+ + \frac{(2 \pi \alpha')^4}{4}\omega_-^2 }\right]^{-1}, 
\end{eqnarray}
where $\omega_+ = (\omega + \bar \omega)$ and $\omega_- = (\omega - \bar \omega)$.

 It is possible to write a supersymmetric version of this action. This can be done by first defining 
$
 K  = D^2[W^a W_{ a}], $ and $   \bar K  = \bar D^2 [\bar W^a_2\bar W_{2a}]. 
$, and then   written  the supersymmetric abelian  Born-Infeld Lagrangian as
  \begin{eqnarray}
  \mathcal{L} &=& D^2  [W^a W_a]_{\theta= 0} 
  + \bar D^2 [\bar W^a \bar W_a]_{\bar \theta = 0}  \nonumber \\ && +
\bar D^2  D^2   [W^a W_{ a}\bar W^b \bar W_{b} \mathcal{B}  (K, \bar K) ]_{\theta = \bar\theta =0 }. 
 \end{eqnarray} 
The constraint  $\mathcal{B} (K , K )$ can be written as \cite{1b}
  \begin{eqnarray}
  B (K , \bar K ) =  \left[1- \frac{(2 \pi\alpha')^2}{2}K_+
  + \sqrt{ 1 + (2 \pi\alpha')^2K_+ + \frac{(2 \pi \alpha')^4}{4}K_-^2 }\right]^{-1},
\end{eqnarray}
where  $K_+ = (K + \bar K)$ and $K_- = (K - \bar K)$.
The    abelian Born-Infeld Lagrangian can be written as 
  \begin{eqnarray}
  \mathcal{L} &=& D^2  [W^a W_a]_{\theta= 0} 
  + \bar D^2 [\bar W^a \bar W_a]_{\bar \theta = 0}  \nonumber \\ && +
\bar D^2  D^2   [W^a W_{ a}\bar W^b \bar W_{b} \mathcal{B}  (K, \bar K) ]_{\theta = \bar\theta =0 }
\nonumber \\ &=& 
 D^2 \bar D^2 [\Box^{-1}   D^2   W^a W_a     +
 + \Box^{-1} \bar D^2    \bar W^a \bar W_a    \nonumber \\
  &&  + W^a W_{ a}\bar W^b \bar W_{b} \mathcal{B}  (K, \bar K) ]_{\theta = \bar\theta =0 },
 \end{eqnarray} 
We can transform this Lagrangian  to the one containing $W_{1a}$ and $W_{2a}$ as follows, 
\begin{eqnarray}
 \mathcal{L} &=& D^2_1 D^2_2 [W^a_1W_{1a} W^a_2W_{2a}\mathcal{B}(K_1, K_2)]_{\theta_1 = \theta_2 =0}\nonumber \\&&
+ D^2_1 [W^a_1W_{1a}]_{\theta_1 =0} + D^2_2
[W^a_2W_{2a}]_{\theta_2 =0},
\end{eqnarray}
where $
 K_2  = D^2_1[W^a_1 W_{1 a}], $ and $   K_2  =   D^2_2 [ W^a_2\bar W_{2a}]. 
$
So, we can write the abelian Born-Infeld Lagrangian  in presence of a boundary as 
\begin{eqnarray}
 \mathcal{L}^{1- 2 -} &=& (D^2_1 - \partial_3) (D^2_2 - \partial_3)
[W^a_1W_{1a} W^a_2W_{2a}\mathcal{B}(K_1, K_2)]_{\theta_1 = \theta_2 =0}
  \nonumber \\&&+ (D^2_1 - \partial_3) [W^a_1W_{1a}]_{\theta_1 =0} + (D^2_2 - \partial_3) [ W^a_2W_{2a}]_{\theta_2 =0}
 , \nonumber \\
 \mathcal{L}^{1- 2 +} &=& (D^2_1 - \partial_3) (D^2_2+ \partial_3)
[ W^a_1W_{1a} W^a_2W_{2a}\mathcal{B}(K_1, K_2)]_{\theta_1 = \theta_2 =0}
 \nonumber \\&& +(D^2_1 - \partial_3) [W^a_1W_{1a}]_{\theta_1 =0} + (D^2_2+ \partial_3)[ W^a_2W_{2a}]_{\theta_2 =0}
 , \nonumber \\
 \mathcal{L}^{1+ 2 -} &=& (D^2_1 + \partial_3) (D^2_2 - \partial_3)
[ W^a_1W_{1a} W^a_2W_{2a}\mathcal{B}(K_1, K_2)]_{\theta_1 = \theta_2 =0}
  \nonumber \\&&+(D^2_1 + \partial_3) [W^a_1W_{1a}]_{\theta_1 =0} + (D^2_2 - \partial_3)[ W^a_2W_{2a}]_{\theta_2 =0}
 , \nonumber \\
 \mathcal{L}^{1+ 2 + } &=& (D^2_1+ \partial_3) (D^2_2 + \partial_3)
[ W^a_1W_{1a} W^a_2W_{2a}\mathcal{B}(K_1, K_2)]_{\theta_1 = \theta_2 =0}
  \nonumber \\&&+  (D^2_1+ \partial_3) [W^a_1W_{1a}]_{\theta_1 =0} + (D^2_2 + \partial_3) [ W^a_2W_{2a}]_{\theta_2 =0}.
\end{eqnarray}
This result can be also be written as
 \begin{eqnarray}
   \mathcal{L}^{1+ 2+} &=&  D_{2+}D_{2-}D_{1+}D_{1-}
 [  W^a_1W_{1a} W^a_2W_{2a}\mathcal{B}(K_1, K_2)]_{\theta_1 = \theta_2=0}
  \nonumber \\&&+ D_{1+}D_{1-} [W^a_1W_{1a}]_{\theta_1 =0} + D_{2+}D_{2-} [ W^a_2W_{2a}]_{\theta_2 =0}
 ,
\nonumber \\
   \mathcal{L}^{1- 2-} &=&  D_{2-}D_{2+}D_{1-}D_{1+}
 [  W^a_1W_{1a} W^a_2W_{2a}\mathcal{B}(K_1, K_2)]_{\theta_1 = \theta_2=0}
  \nonumber \\&&+ D_{1-}D_{1+} [W^a_1W_{1a}]_{\theta_1 =0} + D_{2-}D_{2+} [ W^a_2W_{2a}]_{\theta_2 =0}
 ,
\nonumber \\
   \mathcal{L}^{1+ 2-} &=&  D_{2+}D_{2-}D_{1-}D_{1+}
 [W^a_1W_{1a} W^a_2W_{2a}\mathcal{B}(K_1, K_2)]_{\theta_1 = \theta_2=0}
  \nonumber \\&&+ D_{1-}D_{1+} [W^a_1W_{1a}]_{\theta_1 =0} + D_{2+}D_{2-} [ W^a_2W_{2a}]_{\theta_2 =0}
 ,
\nonumber \\
   \mathcal{L}^{1- 2+} &=& D_{2-}D_{2+}D_{1+}D_{1-}
 [W^a_1W_{1a} W^a_2W_{2a}\mathcal{B}(K_1, K_2) ]_{\theta_1 = \theta_2=0}
  \nonumber \\&&+ D_{1+}D_{1-} [W^a_1W_{1a}]_{\theta_1 =0} + D_{2-}D_{2+} [ W^a_2W_{2a}]_{\theta_2 =0}.
 \end{eqnarray}

 The abelian Born-Infeld Lagrangian can couple to a background dilaton $\phi$ and an axion $C$.
 The supersymmetric version of this action will also require a dilatino field $\lambda_a$. 
To write the action for the system, we define a  complex scalar 
$
  \rho = e^{-\phi} + i C 
$.
We can write
$A = \rho + \theta^a \lambda_a + \theta^2 F.$ and $\bar A = 
\bar \rho + \bar \theta^a \bar \lambda_a + \bar \theta^2 \bar F $. Here  $F$ and 
$\bar F$ are  auxiliary fields. We can also define $\mathcal{A} = A + \bar A $. 
This Lagrangian for this system can now be written as \cite{1b}
  \begin{eqnarray}
  \mathcal{L} &=& D^2  [W^a W_a]_{\theta= 0} 
  + \bar D^2 [\bar W^a \bar W_a]_{\bar \theta = 0}  \nonumber \\ && +
\bar D^2  D^2   [ \mathcal{A} ^2 
W^a W_{ a}\bar W^b \bar W_{b} \mathcal{B}  (K, \bar K,  \mathcal{A})  ]_{\theta = \bar\theta =0 }
\nonumber \\ &=& 
 D^2 \bar D^2 [\Box^{-1}   D^2   W^a W_a     +
 + \Box^{-1} \bar D^2    \bar W^a \bar W_a    \nonumber \\
  &&  + \mathcal{A}^2 
W^a W_{ a}\bar W^b \bar W_{b} \mathcal{B}  (K, \bar K,  \mathcal{A})
 ]_{\theta = \bar\theta =0 }. 
 \end{eqnarray} 
The constraint  $\mathcal{B}  (K, \bar K, \mathcal{A})$ can be written as 
  \begin{eqnarray}
\mathcal{B}  (K, \bar K,  \mathcal{A}) =  \left[1- \frac{(2 \pi\alpha')^2}{2}\mathcal{A}_+
  + \sqrt{ (1 + (2 \pi\alpha')^2 \mathcal{A}_+ +    \frac{(2 \pi \alpha')^4}{4}\mathcal{A}_-^2 }\right]^{-1},
\end{eqnarray}
where  $2\mathcal{A}_+ = (   \mathcal{A}  K +    \mathcal{A} \bar K)$ and $2\mathcal{A}_- = (   \mathcal{A} K 
-    \mathcal{A} \bar K)$. We can again 
transform this Lagrangian  to the one containing $W_{1a}$ and $W_{2a}$ as follows, 
\begin{eqnarray}
 \mathcal{L} &=& D^2_1 D^2_2 [   \mathcal{A}^2
 W^a_1W_{1a} W^a_2W_{2a}\mathcal{B}(K_1, K_2,  \mathcal{A})]_{\theta_1 = \theta_2 =0}\nonumber \\&&
+ D^2_1 [W^a_1W_{1a}]_{\theta_1 =0} + D^2_2
[W^a_2W_{2a}]_{\theta_2 =0}. 
\end{eqnarray}
Here the $\mathcal{A}$ has also been transformed to the superspace coordinates $\theta_1$ and $\theta_2$, 
and the Jacobian of the transformation has been absorbed in the field redefinition. 
Now in presence of a boundary, a Born-Infeld Lagrangian coupled to a  dilaton and an axion is given by 
\begin{eqnarray}
 \mathcal{L}^{1- 2 -} &=& (D^2_1 - \partial_3) (D^2_2 - \partial_3)
[   \mathcal{A} ^2
 W^a_1W_{1a} W^a_2W_{2a}  \nonumber \\&&\times \mathcal{B}(K_1, K_2,  \mathcal{A})]_{\theta_1 = \theta_2 =0}
  \nonumber \\&&+ (D^2_1 - \partial_3) [W^a_1W_{1a}]_{\theta_1 =0} + (D^2_2 - \partial_3) [ W^a_2W_{2a}]_{\theta_2 =0}
 , \nonumber \\
 \mathcal{L}^{1- 2 +} &=& (D^2_1 - \partial_3) (D^2_2+ \partial_3)
[    \mathcal{A} ^2
 W^a_1W_{1a} W^a_2W_{2a}  \nonumber \\&&\times \mathcal{B}(K_1, K_2,  \mathcal{A})]_{\theta_1 = \theta_2 =0}
 \nonumber \\&& +(D^2_1 - \partial_3) [W^a_1W_{1a}]_{\theta_1 =0} + (D^2_2+ \partial_3)[ W^a_2W_{2a}]_{\theta_2 =0}
 , \nonumber \\
 \mathcal{L}^{1+ 2 -} &=& (D^2_1 + \partial_3) (D^2_2 - \partial_3)
[    \mathcal{A} ^2
 W^a_1W_{1a} W^a_2W_{2a}  \nonumber \\&&\times \mathcal{B}(K_1, K_2,  \mathcal{A})]_{\theta_1 = \theta_2 =0}
  \nonumber \\&&+(D^2_1 + \partial_3) [W^a_1W_{1a}]_{\theta_1 =0} + (D^2_2 - \partial_3)[ W^a_2W_{2a}]_{\theta_2 =0}
 , \nonumber \\
 \mathcal{L}^{1+ 2 + } &=& (D^2_1+ \partial_3) (D^2_2 + \partial_3)
[    \mathcal{A} ^2
 W^a_1W_{1a} W^a_2W_{2a}  \nonumber \\&&\times \mathcal{B}(K_1, K_2,  \mathcal{A})]_{\theta_1 = \theta_2 =0}
  \nonumber \\&&+  (D^2_1+ \partial_3) [W^a_1W_{1a}]_{\theta_1 =0} + (D^2_2 + \partial_3) [ W^a_2W_{2a}]_{\theta_2 =0}.
\end{eqnarray}
This result can be also be written as
 \begin{eqnarray}
   \mathcal{L}^{1+ 2+} &=&  D_{2+}D_{2-}D_{1+}D_{1-}
 [    \mathcal{A} ^2
 W^a_1W_{1a} W^a_2W_{2a}  \nonumber \\&&\times \mathcal{B}(K_1, K_2,  \mathcal{A})]_{\theta_1 = \theta_2=0}
  \nonumber \\&&+ D_{1+}D_{1-} [W^a_1W_{1a}]_{\theta_1 =0} + D_{2+}D_{2-} [ W^a_2W_{2a}]_{\theta_2 =0}
 ,
\nonumber \\
   \mathcal{L}^{1- 2-} &=&  D_{2-}D_{2+}D_{1-}D_{1+}
 [    \mathcal{A} ^2
 W^a_1W_{1a} W^a_2W_{2a}  \nonumber \\&&\times \mathcal{B}(K_1, K_2,  \mathcal{A})]_{\theta_1 = \theta_2=0}
  \nonumber \\&&+ D_{1-}D_{1+} [W^a_1W_{1a}]_{\theta_1 =0} + D_{2-}D_{2+} [ W^a_2W_{2a}]_{\theta_2 =0}
 ,
\nonumber \\
   \mathcal{L}^{1+ 2-} &=&  D_{2+}D_{2-}D_{1-}D_{1+}
 [   \mathcal{A} ^2
 W^a_1W_{1a} W^a_2W_{2a}  \nonumber \\&&\times \mathcal{B}(K_1, K_2,  \mathcal{A})]_{\theta_1 = \theta_2=0}
  \nonumber \\&&+ D_{1-}D_{1+} [W^a_1W_{1a}]_{\theta_1 =0} + D_{2+}D_{2-} [ W^a_2W_{2a}]_{\theta_2 =0}
 ,
\nonumber \\
   \mathcal{L}^{1- 2+} &=& D_{2-}D_{2+}D_{1+}D_{1-}
 [    \mathcal{A} ^2
 W^a_1W_{1a} W^a_2W_{2a}  \nonumber \\&&\times \mathcal{B}(K_1, K_2,  \mathcal{A}) ]_{\theta_1 = \theta_2=0}
  \nonumber \\&&+ D_{1+}D_{1-} [W^a_1W_{1a}]_{\theta_1 =0} + D_{2-}D_{2+} [ W^a_2W_{2a}]_{\theta_2 =0}.
 \end{eqnarray}

The abelian Born-Infeld Lagrangian in absence of a  dilaton and axion  can also be written as a 
non-linear sigma model \cite{1b}, 
\begin{eqnarray}
 \mathcal{L} = D^2 [\chi ]_{\theta = 0} 
 +  \bar D^2 [\bar \chi]_{\bar \theta = 0},   
\end{eqnarray}
where 
\begin{eqnarray}
 \chi + \frac{(2 \pi \alpha')^2}{2}\chi \bar D^2 \bar \chi  &=&   \frac{1}{4} W^a W_a,  \nonumber \\
 \bar \chi + \frac{(2 \pi \alpha')^2}{2}\bar \chi   D^2   \chi &=&    \frac{1}{4}\bar  W^a \bar W_a. 
\end{eqnarray}
It is possible to extend this formalism to non-abelian gauge theories. This can be done by 
defining \cite{2b}
\begin{eqnarray}
 \xi +\frac{(2 \pi \alpha')^2}{2} \xi  \bar D^2 ( e^{2V}\bar  \xi  e^{2V})  &=&   
 \frac{1}{4} W^a W_a,  \nonumber \\
  {\bar \xi} + \frac{(2 \pi \alpha')^2}{2}\bar  \xi    D^2   (e^{-2V}  \xi  e^{2V})  &=& 
   \frac{1}{4}\bar  W^a \bar W_a, 
\end{eqnarray}
where  $W^a$ and $\bar W^a$ are field strengths for non-abelian gauge theories. 
Now the non-abelian Born-Infeld Lagrangian can be written as 
\begin{eqnarray}
 \mathcal{L} &=& D^2 [\xi ]_{\theta = 0} 
 +  \bar D^2 [\bar \xi]_{\bar \theta = 0} \nonumber \\
 &=& D^2 \bar D^2 [\Box^{-1} \xi ]_{\theta = 0} 
 +  \bar D^2 D^2 [\Box^{-1} \bar \xi]_{\bar \theta = 0}.
\end{eqnarray}
Now we  define $\tilde \zeta (\theta, \bar \theta)= \Box^{-1} \xi  + \Box^{-1} \bar \xi$, and 
transform it to
\begin{eqnarray}
 \zeta (\theta_1, \theta_2) =
\mathcal{J}  \tilde \zeta (\theta_1, \theta_2) 
\end{eqnarray}
 where $\mathcal{J} $ is the Jacobian for transformation from $\theta, \bar \theta$
 to $\theta_1, \theta_2$. 
 So, we can write the non-abelian Born-Infeld Lagrangian as
 \begin{eqnarray}
 \mathcal{L}  =  D_1^2 D_2^2 [\zeta (\theta_1, \theta_2)]_{\bar \theta = 0}. 
\end{eqnarray}
It is possible to couple this action to matter fields  and write the combined action as 
\begin{eqnarray}
  \mathcal{L}  =  D_1^2 D_2^2 [\zeta (\theta_1, \theta_2) +  \nabla^a \Phi  \bar \nabla_a  \bar  \Phi
 + \mathcal{V} [\Phi, \bar \Phi]]_{\bar \theta = 0}. 
\end{eqnarray}
So, we can write the action for the non-abelian Born-Infeld Lagrangian coupled to matter fields as
\begin{eqnarray}
 \mathcal{L}^{1- 2 -} &=& (D^2_1 - \partial_3) (D^2_2 - \partial_3)
[ \zeta (\theta_1, \theta_2) + \nabla^a \Phi  \bar \nabla_a  \bar  \Phi
  \nonumber \\ && + \mathcal{V} [\Phi, \bar \Phi]]_{\theta_1 = \theta_2 =0}
 , \nonumber \\
 \mathcal{L}^{1- 2 +} &=& (D^2_1 - \partial_3) (D^2_2+ \partial_3)
[ \zeta (\theta_1, \theta_2) +\nabla^a \Phi  \bar \nabla_a  \bar  \Phi
  \nonumber \\ && + \mathcal{V} [\Phi, \bar \Phi]]_{\theta_1 = \theta_2 =0}
 , \nonumber \\
 \mathcal{L}^{1+ 2 -} &=& (D^2_1 + \partial_3) (D^2_2 - \partial_3)
[ \zeta (\theta_1, \theta_2) +\nabla^a \Phi  \bar \nabla_a  \bar  \Phi
  \nonumber \\ && + \mathcal{V} [\Phi, \bar \Phi]]_{\theta_1 = \theta_2 =0}
 , \nonumber \\
 \mathcal{L}^{1+ 2 + } &=& (D^2_1+ \partial_3) (D^2_2 + \partial_3)
[ \zeta (\theta_1, \theta_2) +\nabla^a \Phi  \bar \nabla_a  \bar  \Phi
  \nonumber \\ && + \mathcal{V} [\Phi, \bar \Phi]]_{\theta_1 = \theta_2 =0}
 .
\end{eqnarray}
This result can be also be written as
 \begin{eqnarray}
   \mathcal{L}^{1+ 2+} &=&  D_{2+}D_{2-}D_{1+}D_{1-}
 [ \zeta (\theta_1, \theta_2) + \nabla^a \Phi  \bar \nabla_a  \bar  \Phi
  \nonumber \\ && + \mathcal{V} [\Phi, \bar \Phi]]_{\theta_1 = \theta_2=0}
 ,
\nonumber \\
   \mathcal{L}^{1- 2-} &=&  D_{2-}D_{2+}D_{1-}D_{1+}
 [ \zeta (\theta_1, \theta_2) + \nabla^a \Phi  \bar \nabla_a  \bar  \Phi
  \nonumber \\ && + \mathcal{V} [\Phi, \bar \Phi]]_{\theta_1 = \theta_2=0},
\nonumber \\
   \mathcal{L}^{1+ 2-} &=&  D_{2+}D_{2-}D_{1-}D_{1+}
 [ \zeta (\theta_1, \theta_2) + \nabla^a \Phi  \bar \nabla_a  \bar  \Phi
  \nonumber \\ && + \mathcal{V} [\Phi, \bar \Phi]]_{\theta_1 = \theta_2=0}
 ,
\nonumber \\
   \mathcal{L}^{1- 2+} &=& D_{2-}D_{2+}D_{1+}D_{1-}
 [ \zeta (\theta_1, \theta_2) +\nabla^a \Phi  \bar \nabla_a  \bar  \Phi
 \nonumber \\ &&  + \mathcal{V} [\Phi, \bar \Phi] ]_{\theta_1 = \theta_2=0}.
 \end{eqnarray}
 Thus, we have been able to analyse the non-abelian 
 Born-Infeld Lagrangian coupled to matter fields, in presence of a boundary. 
 This Lagrangian  also preserves only half the supersymmetry of the original Lagrangian.

 \section{Conclusion}\label{hgfb1}
In this paper, we have analysed the restoration of half the supersymmetry for a four dimensional theory
in $\mathcal{N} =1$ superspace
formalism, on manifolds with a boundary. We first use the fact that a total derivative term is obtained from the
supersymmetric variation of a Lagrangian for a four dimensional theory with  $\mathcal{N} =1$ supersymmetry. 
This total derivative term vanishes in absence of a boundary.
 However, in presence of a boundary, this total derivative term generates a boundary term, which breaks half the
 supersymmetry of the original theory. However, half of the original supersymmetry
 can be preserved by adding new boundary terms to the original Lagrangian.
The supersymmetric variation of these new boundary terms exactly canceled the boundary terms generated by the
supersymmetric  transformation
of the original bulk Lagrangian.  We explicitly constructed such boundary terms for the four dimensional theory
with $\mathcal{N} =1$ supersymmetry. We also related the bulk supercharges to the boundary supercharges.
The bulk supercharges behaved as two independent
supercharges on the boundary. However, the inclusion of the new boundary terms only preserved the supersymmetry
only with respect to one of these projections. Thus, it was demonstrated that only half of the supersymmetry of 
the original theory
was preserved. This analysis was done using the real superfields, and the Jacobian of transformation was absorbed 
in field redefinitions. We finally applied our results to the super-Yang-Mills theory. We explicitly constructed
the Lagrangian which
preserves half the supersymmetry of the original theory. 
We also study the Born-Infeld Lagrangian in presence of a boundary. We study the coupling of the Born-Infeld 
Lagrangian to a dilaton and an axion field. We also study the non-abelian Born-Infeld action. We demonstrate 
that the Born-Infeld Lagrangian preserves half the supersymmetry of the original theory, in presence of a boundary.

It is possible to generalize this present analysis to theories which have higher amount of supersymmetry.
In fact, the
  analysis of this present paper can be used for analysing various aspects of the $AdS/CFT$ correspondence 
  \cite{1g}-\cite{1c}.
 This is because according to the $AdS/CFT$ correspondence
  type IIB string theory on $AdS_5 \times S^5$ is dual to the $\mathcal{N}=4$ super-Yang-Mills theory on  its conformal  boundary.
  Thus,  the theory which describes the low energy limit of the action for a stack  of  D3-branes on $AdS_5 \times S^5$
  is the  $\mathcal{N} = 4$ super-Yang-Mills action
  with the gauge group $U(N)$. The
 four would-volume coordinates of the D3-branes become the Minkowski coordinates, 
 and six transverse coordinates to the  D3-branes
 give rise to
  the six gauge valued scalar fields of $\mathcal{N} =4$  super-Yang-Mills theory. This theory also contains
  eight gauge valued fermions, and a gauge field.  It would be interesting to analyse $\mathcal{N}=4$ 
  super-Yang-Mills theory
  in presence of a boundary.  It is again expected that the $\mathcal{N}=4$ super-Yang-Mills theory
  in presence of a boundary will preserve only half the supersymmetry of the original theory.

It has been demonstrated that using the Horava-Witten  theory,    one of the low energy limits of the heterotic string theory can
be obtained from the  eleven dimensional supergravity in presence of a  boundary  \cite{z1}-\cite{1z}.
In this construction,
it has been  possible to obtain  a unification of gauge and gravitational couplings.
It would be interesting to analyse the connection between the Horava-Witten  theory and the boundary 
supersymmetry discussed in this paper.
In order to do that, it might be interesting to first generalize the results of this paper to
five dimensions. This is because,
  motivated by Horava-Witten  theory,
  a five dimensional globally supersymmetric Yang-Mills theory coupled to a four dimensional hypermultiplet
on the boundary has  been constructed  \cite{z2}.

It may be noted that in Randall-Sundrum  models our four dimensional universe is thought to be located
on a three-brane in a five-dimensional spacetime with negative cosmological constant \cite{q1}-\cite{q2}.
These models provide a geometrical solution to the electroweak hierarchy problem.
A supersymmetric generalization of such models have also been analysed in \cite{p1}-\cite{p2}.
In fact, it has been argued that for the supersymmetric Randall-Sundrum models to be consistent, the issue of supersymmetric boundary
conditions has  to be analysed \cite{p4}. We are hoping to generalize such a procedure to help constructing
supersymmetric Randall-Sundrum models.

\end{document}